\documentclass[showpacs,twocolumn,prl]{revtex4}
\usepackage{amsmath,graphicx}
\usepackage{color,ulem}

\definecolor{darkred}{rgb}{0.90,0,0}
\definecolor{darkgreen}{rgb}{0,0.60,.2}
\definecolor{darkblue}{rgb}{0,0,1}
\definecolor{grey}{cmyk}{0,0,0,0.25}
\definecolor{orange}{cmyk}{0,0.6,0.8,0}

\newcommand{\DM}[1]{\textcolor{black}{#1}}
\newcommand{\DLM}[1]{\textcolor{black}{#1}}
\newcommand{\DLMB}[1]{\textcolor{black}{#1}}

\begin{document}

\title {Transport and elastic scattering times  as probes of the nature of  impurity scattering in single and bilayer graphene}
\author{M. Monteverde,$^{1}$ C. Ojeda-Aristizabal,$^{1}$ R. Weil,$^{1}$ K. Bennaceur,$^{2}$ M. Ferrier,$^{1}$ S. Gu\'eron,$^{1}$ C. Glattli,$^{2}$ H. Bouchiat,$^{1}$ J. N.  Fuchs,$^{1}$  
and D. L. Maslov$^{1,3}$}
\affiliation{$^{1}$LPS, Univ. Paris-Sud, CNRS, UMR 8502, F-91405 Orsay Cedex, France\\$^{2}$Service de Physique de l'Etat Condensée/IRAMIS/DSM (CNRS URA 2464), CEA Saclay, F-91191 Gif-sur-Yvette, France\\$^{3}$Department of Physics, University of Florida,
 Gainesville, FL 32611, USA}
\pacs{63.22.Np, 73.23.Hk, 73.21.Ac} 
\begin{abstract}
Both transport  $\tau_{tr}$  and elastic $\tau_{e}$ scattering times   are  experimentally determined from the carrier density dependence of the magnetoconductance of monolayer and bilayer graphene.  Both times and their dependences on  
carrier density  are  found to be very different in the monolayer and the bilayer. However, their ratio $\tau_{tr}/\tau_{e} $ is found to be close to $1.8 $ in both systems and  nearly independent of the carrier density. These measurements give insight on the nature (neutral or charged) and range of  the scatterers. Comparison with theoretical predictions 
suggests that the main scattering mechanism  in our samples is due to strong \DLM{(resonant})
scatterers of a range \DLM{shorter} than the Fermi wavelength;  likely candidates being vacancies, voids, ad-atoms or short-range ripples.

\end{abstract}

\maketitle

Since the discovery of the fascinating electronic properties of graphene  \cite{revueguinea} due to  its  electronic spectrum  with linear  dispersion and a perfect  electron-hole symmetry at the  Fermi level \cite{bandstructurewallace},  the nature of defects  has been shown to play an essential role in determining the carrier density 
($n_c$) dependence of the conductance.
 The wavevector  and energy dependences of the impurity potential are known to determine the  characteristic scattering times of the  carriers. It is important to distinguish the transport time $\tau_{tr}$, \DM{which governs the current relaxation and enters}  
the Drude conductivity ($\sigma$), from the elastic scattering time $\tau_{e}$\DM{,} which is the lifetime of a plane wave state \cite{gilles}. 
\DM{Since $\tau_{tr}$ and $\tau_{e}$} involve different angular integrals of the
\DM{differential} cross section\DM{,}
they differ as soon as the  Fourier components of the potential  depend on  \DM{the} wavevector $q$.  A large ratio  $\tau_{tr}/\tau_{e}$   indicates that \DM{scattering is predominantly in the forward direction,}  so that transport
 is not \DM{affected} much by this type of scattering. This is the case in 2D electron gases 
 (2D\DM{E}G) confined  \DM{to} GaAs/GaAlAs  heterojunctions 
 \DLM{with} 
 the scattering potential 
  produced by 
 remote charged Si donors   \cite{tauasga},  \DLM{where} $\tau_{tr}/\tau_{e}$ is found to be larger than 10. 

 The nature of the main scattering mechanism limiting the carrier mobility in graphene is still subject to controversy. It has indeed been shown \cite{Shon, Aleiner, Mirlin} that  "white noise" \DLM{($q$ independent)} scattering 
  leads to  a weak (logarithmic) dependence of 
 $\sigma(n_c)$, in  contradiction with experiments which typically find a linear increase. In contrast, scattering on charged impurities  originates from a $q$ dependent  screened Coulomb  potential  described in the Thomas Fermi approximation \cite{Mcdonald, dasarmabilayer,fuhrer}. This leads to a  linear  $\sigma(n_c)$ both for a monolayer (ML) and a  moderately doped bilayer (BL). Recent experiments performed  to probe this question measured  the change  in  $\sigma$ upon immersion of graphene samples in high\DM{-$K$} dielectric media. Their conclusions differ \cite{geim}.  
  Alternate explanations  involve  resonant scattering centers  with a large energy mismatch with the Fermi energy of carriers \cite{Mirlin,stauber}.
\begin{figure}
\begin{center}
\includegraphics[clip=true,width=8cm]{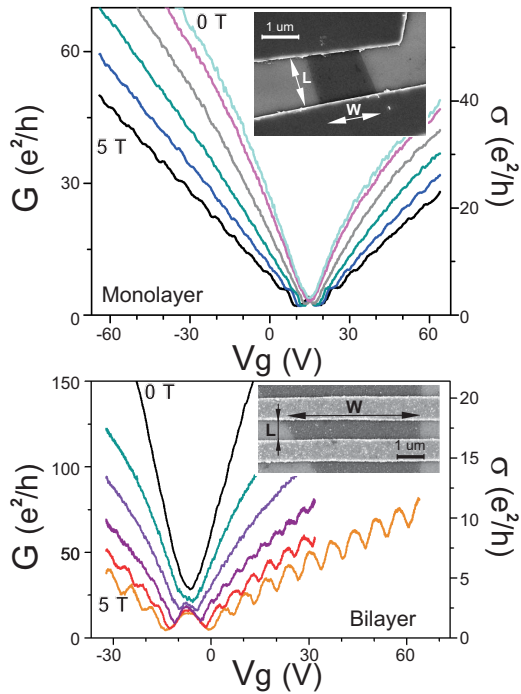}
\caption{Gate voltage  dependence of the conductance  at several magnetic fields, 1 $T$ apart.  The contact resistances have been subtracted. Top panel: monolayer A. Bottom panel: bilayer B. Inset: electron micrographs of the samples. \label{fig1}}
\end{center}
\end{figure}

In order to gain insight into the scattering mechanism in graphene, we have extracted  $\tau_e$ and $\tau_{tr}$ from 
magnetotransport in monolayer and bilayer  graphene samples. In high magnetic field, when the cyclotron frequency is larger than $1/\tau_e$, the magneto-conductivity exhibits Shubnikov de Haas (ShdH) oscillations related to the formation of Landau levels.  The  broadening of these levels at low temperature yields $\tau_e$, while the  low field quadratic magneto-conductivity  yields $\tau_{tr}$ .
  
The samples were  fabricated by exfoliation of  natural graphite flakes and deposition on a doped silicon substrate with a 285 nm thick oxide. 
The carrier density can be tuned from electrons to holes through the charge neutrality point by applying a  voltage on the backgate. The ML and BL samples were identified using Raman spectroscopy. The electrodes were fabricated by electron beam lithography and either sputter deposition of  40 nm thick palladium (samples A and B), or Joule evaporation of a bilayer 5nmTi/70nm Au (other samples C,D and E).  We  mostly discuss samples A and B, a ML and a BL of respective dimensions $W = 1.6\;\mu$m $L=1.3\;\mu$m  and $W = 4.8\;\mu$m, $L=0.7\;\mu$m, \DLM{where} $L$ is the distance between the voltage probes covering nearly the  entire sample  width $W$ (see Fig.~1). The contact resistances were measured to be 20 $\Omega$ for the BL and calculated to be 200 $\Omega$ for the ML, and were subtracted. The gate voltage $V_g$ dependence of $\sigma$ is shown for both samples for a range of
magnetic fields in Fig.~1. At zero field, one observes a   slightly sub-linear increase of the conductance on both sides of its minimum at the neutrality point. The mobility varies between 
3000 and 5000 cm$^2$ V$^{-1}$ s$^{-1}$ for the ML,   3000 and 6000 cm$^2$ V$^{-1}$ s$^{-1}$ for the BL. Above $2$T,    \DLM{steps in the conductance of the ML occur near quantized values  $4(n+1/2)e^2/h$, as expected}. 
The oscillations in the BL (with a maximum of conductance at the neutrality point)  look more unusual
but  can be understood given the aspect ratio of the sample (see below). 

We now describe how we extract  $\tau_{tr}$ and $\tau_e$ from the magnetoresistance (MR)(see Fig.~2). The  two-terminal MR  results from    mixing of  the  diagonal ($\rho _{xx}$)  and \DM{off}-diagonal ($\rho _{xy}$) components of the resitivity tensor \cite{levitov,marcus}. The degree of mixing depends on  the aspect ratio of the sample. For a square geometry, close to that of the monolayer, $R(B)= \left [ \rho _{xx}^2 + \rho _{xy}^2 \right]^{1/2}$;   in a short wide sample such as the bilayer $R(B)=(L/W)\left [ \rho _{xx}^2 + \rho _{xy}^2 \right]/ \rho _{xx}$.
Intermediate geometries can be calculated following the model developed in \cite{levitov}.
It is then possible to reconstruct the complete MR from the expressions of the resistivity tensor \cite{sdh} valid in the limit of moderate magnetic field  where ShdH oscillations can be approximated by their first harmonics:
\begin{equation}
\begin{array}{l}
\delta \rho_{xx}(B)/\rho_0 = 4D_T \exp  \left[- \frac{\pi}{\omega_c\tau_{e}}\right] \cos \left[ \frac{j\pi E_F}{\hbar \omega_c }-\phi \right] \\
\rho_{xy}(B) =  \rho_0 \omega_c\tau_{tr} -\delta \rho_{xx}(B)/2\omega_c\tau_{tr},
\end{array}
\label{hf}
\end{equation}
where  $\rho_0 = 1/\sigma$ is the zero-field resistivity  and  $\omega_ c=  e  B /m^* $ is the cyclotron frequency,    $m^* =\hbar k_F/v_F$ is the cyclotron mass which  depends explicitly on the Fermi wave vector $k_F$ for the ML (constant Fermi velocity). On the other hand,  the bilayer's dispersion relation is parabolic at low energy and  $m^*$ can  be approximated  by the effective mass $m_{eff} = 0.035 m_e$, nearly independent  of   the carrier density in the  range of $V_g$ explored where $ |E_F| \leq 80$ meV is smaller by a factor 5 than the energy band splitting \cite{revueguinea}. This value of $m_{eff}$ is confirmed   by the analysis of  the temperature dependence of  ShdH oscillations \cite{mass}. 
 The phase $\phi$, either $\pi$ or $2\pi$, and the parameter $j$, either 1 or 2, depend on the nature of the  sample (ML or BL).   The Fermi energy $E_F$ is $\hbar k_F v_F$ for the monolayer and $\hbar^2 k_F^2/(2m_{eff})$ for the bilayer.
 The prefactor $D_T=\gamma / \sinh(\gamma)$ with $\gamma= 2 \pi^2 k_B T/\hbar \omega_c$ describes the thermal damping of the oscillations. 
 
To analyze the data  we first deduce  $k_F$ from the periodicity of the ShdH oscillations function of $1/B$. We think that this determination 
is more reliable close to the neutrality point  where the sample is possibly inhomogeneous than the estimation of $n_c  = k_F^2/\pi$ from the gate voltage and the capacitance between the doped silicon substrate and the graphene sample \cite{fuhrer2}. Knowing $k_F$ we then determine $\tau_{tr}$ from the  low field quadratic low field magnetoresistance which is found to be independent of temperature between  1 and 4 K:
\begin{equation}
	R(B)-R(0) =\frac{h}{2e^2}\frac{L}{W} \frac{1}{k_F v_F\tau_{tr}} \alpha_g (\omega_ c \tau_{tr})^2
	\label{lf}
\end{equation}
We have used the relation $\sigma = \rho_0 ^{-1}=(2e^2/h) k_F v_F\tau_{tr}$. The dimensionless coefficient $\alpha_g$,  which depends on the aspect ratio of the sample is determined  numerically  following \cite{levitov} and the experimental values of $W$and $L$. It is found equal to $0.53\pm 0.01$ and $0.84 \pm 0.02$ for our ML and BL samples A and B, respectively. It is important to note that this determination of $\tau_{tr}$ is independent of any assumption of the contact resistance on our  two terminal samples. We  finally extract $\tau_e$ from the  damping of the  first harmonic of  ShdH oscillations in the resistivity tensor  in  $ \exp(-\beta/B)$ where  $\beta= \pi \hbar k_F/ e v_F\tau_e$, see  Eq.\ref{hf}.

 \begin{figure}
\begin{center}
\includegraphics[clip=true,width=8cm]{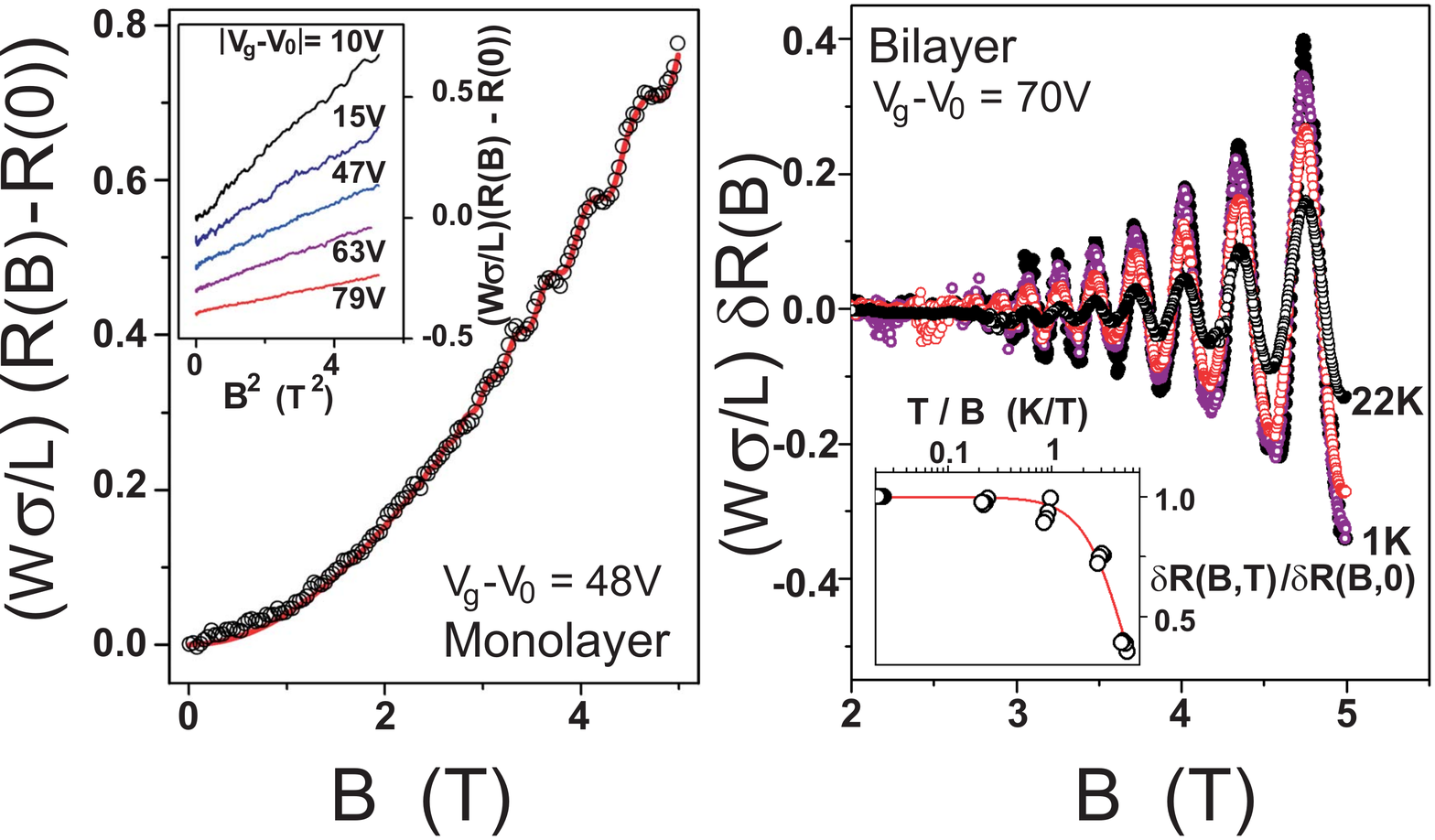}
\caption{Analysis of the magnetoresistance. 
Left  panel: Magnetoresistance of  monolayer sample A.
Dots: experimental points at $T=1$K; Continuous line: fit according to 
Eqs.~(\ref{lf}) and (\ref{hf}). Inset: $B^2$ dependence of the low-field magnetoresistance for different gate voltages (Curves shifted along the Y axis for clarity). $\tau_{tr}$ is extracted from the slopes of these curves according to Eq.~(\ref{lf}). 
Notice that the slope increases in the vicinity of the Dirac point reflecting the divergence of the  inverse effective mass.
Right panel: 
\DLM{ShdH oscillations of the longitudinal component of the  resistivity in bilayer sample B for different temperatures after subtraction of the quadratic backround}. \DLMB{ The Fermi wavevector $k_F$ and the elastic time $\tau_e$  are deduced from the  period and the decay of the oscillations with $1/B$ at low temperature}.  \DLM{ Inset: Temperature dependence of the oscillations amplitude normalized to
$T=0$. Solid line:  fit according to the Lifshitz-Kosevich formula 
$D_T=\gamma/ \sinh(\gamma)$ with $\gamma= 2 \pi^2 k_B T/\hbar \omega_c$ \cite{sdh}. The effective mass 
determined 
from this fit
is $m_{eff}=0.035 \pm 0.002  m_e$ in the whole range of gate voltage investigated. 
}}     

\label{fig2}
\end{center}
\end{figure}

 The $k_F$  dependences of  $\tau_{tr}$ and $\tau_{tr}/\tau_e$ \DM{are}   \DM{shown}  in Fig.~3 for  samples A and B  as well as three other  ML samples, consisting of  another   two-terminal sample C (very similar to A) and two multi-terminal samples  (D and E) with Hall-bar geometry   (see \cite{Suppl} for more details). 
 We observe different behaviors for the ML samples, where $\tau_{tr}$ has a minimum at the CNP, and the BL, where it has a   maximum. In all cases, despite  rather large variations of $\tau_{tr}$\DM{,} $\tau_{tr}/\tau_e$ is nearly independent of $k_F$. It is  equal to $1.7 \pm0.3$  for the monolayers A,C,E and to $1.8\pm0.2$ for the bilayer in the whole range  explored,  which corresponds to $n_c$ between  $1.5\times 10^{11}$ and $5\times 10^{12} \mathrm{cm}^{-2}$. That  $\tau_{tr}/\tau_e$   is of the order but smaller than 2 indicates that the typical size of the scatterers does not exceed the Fermi wavelength. We note however that sample D exhibits a value of $\tau_{tr}/\tau_e $ at high electron doping which is larger than 2 ($\simeq 2.4$).  The area of this sample  ($12 \mu m^2$)  is much larger than the area ($\simeq 1\mu m^2$) of all the other samples A, B, C and E.  We suspect that this large sample contains more   spatial inhomogenities than the other smaller samples which could explain a reduced value of $\tau_e$  .


Finally\DM{,}  it is also possible to fit $\sigma(V_g)$ at 5T depicted in Fig.1  using \DM{the value} of  $\tau_e$ \DM{determined as} described above.
Taking into account the geometry of  samples A and B, following \cite{levitov}, one can relate the contributions of the $n^{\mathrm{th}}$ Landau level to the conductivity tensor, $\delta_n \sigma _{xx}$ and $\delta_n \sigma _{xy}$ within the semi-circular model.  
\begin{equation} 
\begin{array}{l}
\delta_n \sigma _{xx} \sim \exp -ln2\left[(\nu -(\nu_n +\nu_{n+1})/2 )/\Gamma_{\nu}\right]^2\\
\delta_n (\sigma_{xx})^2 +(\delta_n \sigma_{xy} - \sigma^0_{xy,n})(\delta_n \sigma_{xy} - \sigma^0_{xy,n+1})= 0.
\end{array}
\label{levitov}
\end{equation}
\DM{Here, $\nu_{n}$ is the filling factor of the $n^{\mathrm{th}}$ level, $\nu$ is the filling factor in between the $n^{\mathrm{th}}$ and $(n+1)^{\mathrm{th}}$ levels}, and $\sigma^0_{xy,n}$ is the quantized Hall conductivity at the $n^{\mathrm{th}}$ plateau
\DM{equal to} $4(n+1/2)e^2/h$ and $4n e^2/h$  for the ML and BL\DM{,} respectively. 
 The   width of  Landau levels function of filling factor, $\Gamma_\nu $, and energy $\Gamma_E = \hbar\sqrt{2\omega_c /\pi\tau_e}$\cite{ando}, are related via: 
\begin{equation} 
\begin{array}{l}
\displaystyle \Gamma_\nu  = \Gamma_E \frac{2}{\hbar v_F}\sqrt{\frac {\nu_n \Phi_0}{\pi B}}. \mbox{     }
\end{array}
\label{lambda}
\end{equation}
 \begin{figure} 
 \begin{center}
\includegraphics[clip=true,width=8cm]{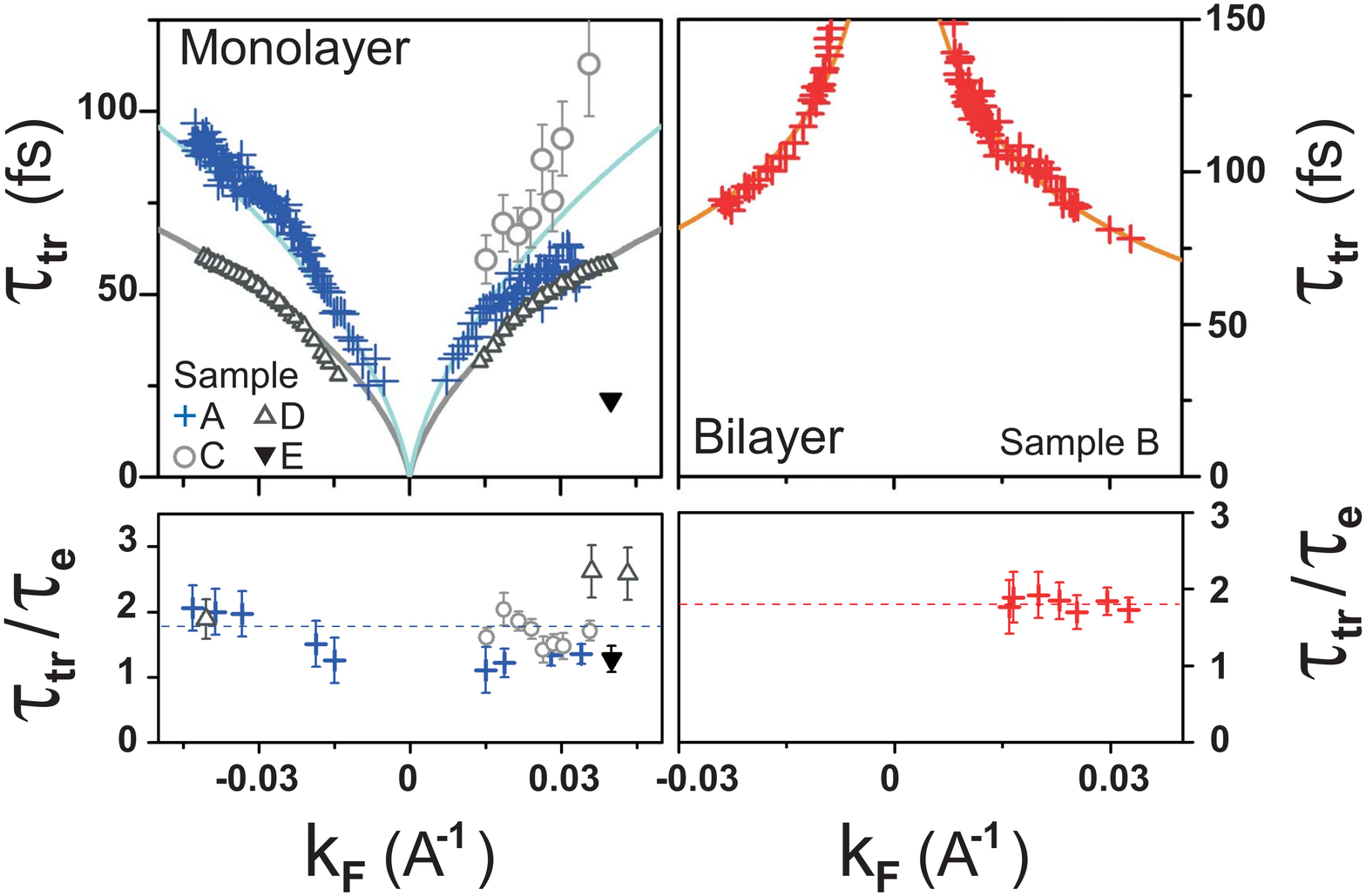}
\caption{ $k_F$ dependence of $\tau_{tr}$ and $\tau_{tr}/\tau_e$ ratio.  Left panel: monolayers  A, C, D and E. Right panel: bilayer B.  The   continuous lines are the fits for samples A,  B and D according to the resonant impurity model, Eq.~\ref{log}. For samples A  B  and C (two terminal configuration)  $\tau_{tr}$ was extracted from the  low field magnetoresistance (crosses)  whereas it was extracted  from the zero field conductivity for samples D and E. Positive/negative values of $k_F$ correspond to electron/hole doping. Lower panels: ratio  $\tau_{tr}/\tau_e$ where $\tau_e$ is  deduced from the fit of the low temperature  decay of the ShdH oscillations. Dotted lines figure the average value $\tau_{tr}/\tau_e = 1.8 $. Interestingly, although the mobilities and accordingly $\tau_{tr}$ vary  substantially from one sample to the other (from 5000 to 800 cm$^2$/Vs from samples A to E) the ratio $\tau_{tr}/\tau_e$ is similar for all samples.}
\label{fig3}
\end{center}
\end{figure} 
Good agreement between the experimental data and the \DM{two-terminal} $G(V_g)$ of the ML, calculated from the conductivity tensor \cite{levitov, marcus}\DM{,} is obtained taking the filling factor dependence  of $\Gamma_\nu$ \DM{from} Eq.~(\ref{lambda}),  see Fig.~4.   
 For the BL,   we had to modify  the semi-circular  relation in a similar way as done in  Ref.~\cite{marcus}. 

We now compare our results on $\tau_e$ and $\tau_{tr}$ to recent theoretical predictions. We first consider scattering on charged impurities \cite{Mcdonald,dasarmabilayer}. In particular, the question of the difference  between $\tau_e$ and $\tau_{tr}$ has been addressed for  a graphene monolayer \cite{sarma08}.  The minimum value of the ratio $\tau_{tr}/\tau_e$ is obtained when the impurities are located close to the graphene foil, in which case it is expected to be   2 -- as a result of the absence of backscattering -- and independent of $n_c$. This is a bit larger than the measured ratio for most samples. Screened charged impurities are characterized by a screening radius $1/q_{sc}$, which in the Thomas-Fermi approximation, is given by $1/q_{TF}\equiv \pi \epsilon \hbar v_F/e^2 k_F$, where $\epsilon$ is the appropriate dielectric constant. In the Born approximation, the transport time is $\tau_{tr}\propto q_{sc}^2 v_F/k_F$. For a monolayer,  $q_{TF}/k_F$ is a constant $\simeq 3$ and both $\tau_{tr}$ and $\tau_e$ are then expected to increase as $k_F$, which is not what we observe in Fig. 3 where the increase is sublinear. The disagreement is even stronger for a bilayer, where the ratio $q_{TF}/k_F \propto 1/k_F$ varies between 3 at high doping  and 12 close to the neutrality point. The transport time is then expected   to vary linearly with   $n_c$, if the screening radius is estimated as $\sim 1/k_F$, or to be independent of $k_F$ if estimated as $\sim 1/q_{TF}\ll 1/k_F$ \cite{dasarmabilayer}, neither of which agrees with our data, see Fig. 3.

 An alternative explanation is resonant scattering resulting from vacancies or any other kind of impurities of 
 range $R$ such that \DLM{$a \lesssim R \ll 1/k_F$}, where $a$ is the carbon-carbon distance,
 and with a large potential energy \cite{Mirlin,stauber}.   It is characterized by a transport cross section
\begin{equation}
	A_{tr} \simeq \frac{\pi^2}{k_F \ln^2(k_FR)}.
	\label{log}
\end{equation}
The resulting transport time $\tau_{tr} =1/(n_iv_F A_{tr})$ ($n_i$ is the concentration of impurities) 
leads to a conductance  increasing as $n_c$  with logarithmic corrections for both the ML and BL. In both cases, our extracted $\tau_{tr}(k_F)$  (see Fig.~3) are compatible with the  square logarithmic dependence of 
 \DM{Eq.~}(\ref{log}). 
 It is \DM{also} possible to estimate the range of the impurity potential $ 0.5\AA \leq R \leq 2.5 \AA$ and 
 the concentration of impurities  $n_i=\DM{(}8 \pm 2\DM{)\times}10^{11}\mathrm{cm} ^{-2}$\DM{,} which \DM{turns out to be}
 identical for sample A and B.   This is of the order of the minimum value of the carrier density  $n_{min}= 1.5\times 10^{11} \mathrm{cm}^{-2}$,  extracted from the experiment.
 It is also interesting to note that the  minimum conductivity expected for this resonant impurity model, $\sigma _{min}= (2 e^2/\pi h)(n_{min}/n_i) ln^2(R \sqrt{\pi n_{min}})= 3.7 e^2/h$  and $4.5 e^2/h $ for the ML and  the BL\DM{,} respectively, are similar to the observed experimental values which are  3.3 and 4.1 $e^2/h$. This analysis also  corroborates our results  on the ratio $\tau_{tr}/\tau_{e}$ indicating  scatterers with a range  smaller than the Fermi wavelength (but possibly of the order if or slightly larger than the lattice spacing).  Whereas the resonant character is not essential for the validity of Eq.~(\ref{log}) for  massive carriers  (corresponding to the  bilayer) \cite{adhikari}, it  has been shown that it is essential for massless carriers in the monolayer \cite{novikov}. This  resonant-like character,  
 \DLMB{although not straightforward},  has been demonstrated in the case of  scattering centers  created by vacancies in graphene  over a wide range of Fermi energies \cite{basko}. As shown in detail in \cite{Suppl}, it is not necessary  to \DM{fine-}tune 
 $k_F$ to obtain the $\ln^2$  dependence \DM{in} Eq. (\ref{log}). 
 
 In conclusion, our results  indicate that the main scattering mechanism  in our graphene samples 
\DM{is} due to strong \DM{ neutral defects,} with a range shorter than the Fermi wavelength and possibly of the order of  $a$, 
 \DLM{inducing resonant (but not unitary) scattering. Likely candidates are} vacancies\DM{,} as observed recently in transmission electron microscopy \cite{zettl}, voids, ad-atoms or short range ripples as suggested in \cite{katsnelson08}. This does not exclude the presence of  long range charged impurities responsible for electron hole puddles but their contribution to the scattering rates $1/\tau_{tr}$ and $1/\tau_{e}$ appears to be negligible in all the samples investigated.

We thank  R. Deblock, M. Goerbig, G. Montambaux, A. Kasumov and A. Mirlin for fruitful discussions.  This work was supported by the
EU-STREP program HYSWITCH, and the "CEE MEST CT 2004 514307 EMERGENT CONDMATPHYS Orsay" grant.

\begin{figure} 
 \begin{center}
\includegraphics[clip=true,width=8cm]{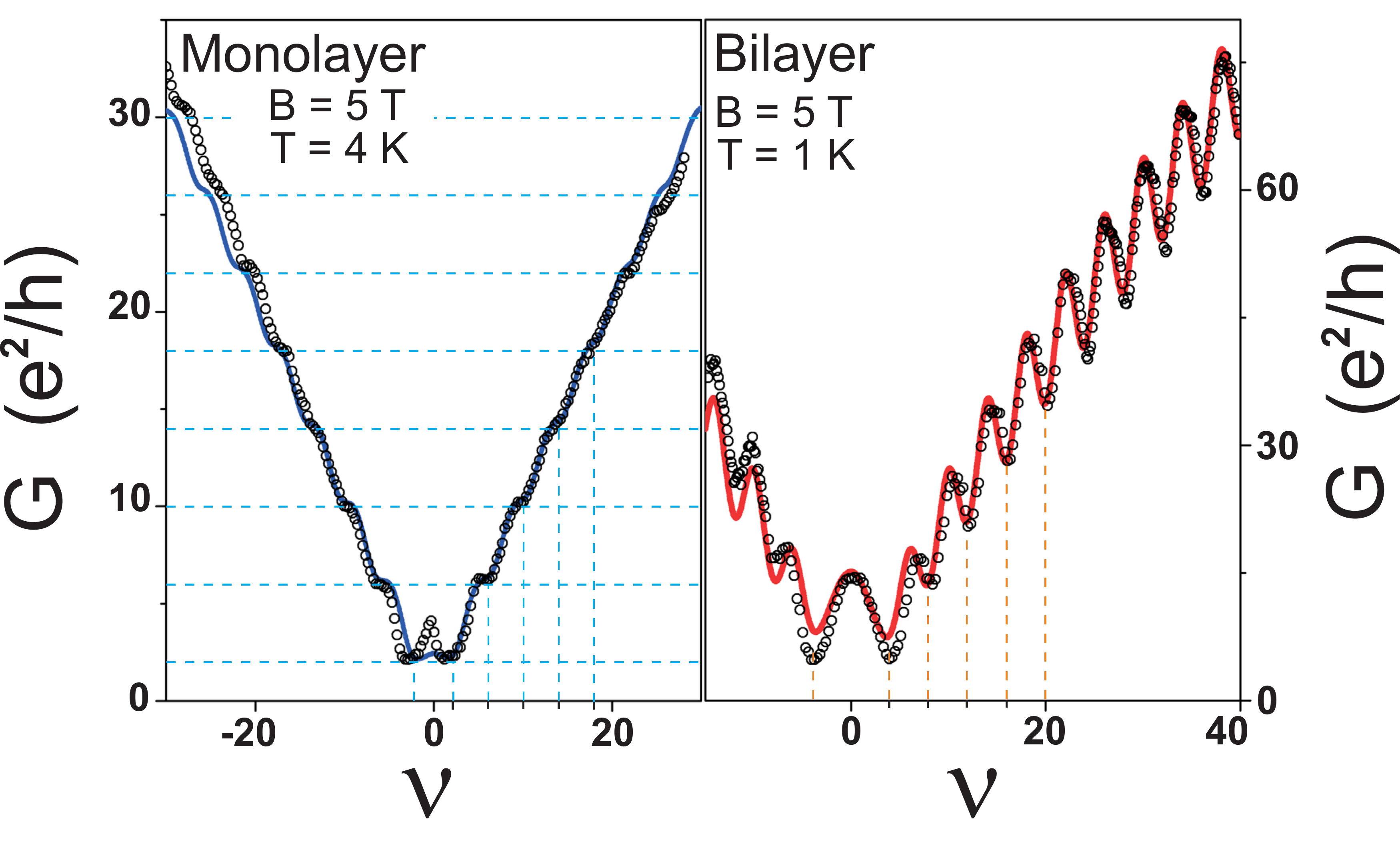}
\caption{Comparison of $G(\nu)$ at 5T for  samples A and B with the  expression of the conductance derived in Ref.~\cite{levitov},taking the  aspect ratio $L/W$ of each sample and  using  Eqs. (\ref{levitov}) and (\ref{lambda})  with $\tau_e(k_F)$  determined above.  The dashed  vertical lines indicate  the positions of $\nu_n$ which, as expected,  are different for the mono $\nu_n =\pm 4(n+1/2)$ and the bilayer $\nu_n =\pm 4n$ . The conductance quantization is well obeyed for the monolayer but  not for the bilayer. This is well explained  by the aspect ratio of the  bilayer sample \cite{marcus, levitov}.}
\label{fig4}
\end{center}
\end{figure}


\begin{thebibliography}{99}


\bibitem{revueguinea} A. H. Castro Neto et al.,
Rev. Mod. Phys. {\bf 81}, 109 (2009).
\bibitem{bandstructurewallace}P. R. Wallace, Phys. Rev. {\bf 71},622 (1947)
\bibitem{gilles}E. Akkermans and G. Montambaux, {\it Mesoscopic Physics with electrons and photons},Cambdrige University Press, (2007).     
\bibitem{tauasga}P. T. Coleridge, Phys. Rev. B{\bf 44}, 3793 (1991).
\bibitem{Shon}N. Shon and T. Ando,  J. Phys. Soc. Japan {\bf 67}, 2421 (1998).
\bibitem{Aleiner} I. L. Aleiner and K. B. Efetov, \prl \textbf{97}, 236801 (2006). 
\bibitem{Mirlin} P. M. Ostrovsky, I. V. Gornyi, and A. D. Mirlin, \prb\textbf{74}, 235443 (2006). 
\bibitem{Mcdonald} K. Nomura, A. H. MacDonald, Phys. Rev. Lett. {\bf 96},
256602 (2006); T. Ando, J. Phys. Soc. Japan {\bf 75}, 074716 (2006);   
\bibitem{dasarmabilayer}S. Adam and S. das Sarma, Phys. Rev. B{\bf 77},115436 (2008). 
\bibitem{fuhrer}S. Adam, S. Cho, M. Fuhrer and S. Das Sarma  Phys. Rev. Lett. 
\DLM{\textbf{101}, 046404 (2008)}.
\bibitem{geim} C. Jang et al.,  Phys. Rev. Lett. {\textbf 101}, 146805(2008),
L. A. Ponomarenko et al.,  Phys. Rev. Lett.
\textbf 
{102}, 206603(2009).
\bibitem{stauber}
 M. I. Katsnelson and  K. S. Novoselov, Solid State Commun. {\bf 143}, 3 (2007), T. Stauber, N. M. T. Peres and F. Guinea, Phys.Rev.{\bf B 76}, 1120 (2007).
\bibitem{fuhrer2}S. Cho and M. S. Fuhrer, Phys. Rev. {\bf B 77}, 081402 (2008).
\bibitem{levitov}D. A. Abanin and L. S. Levitov, Phys. Rev. B{\bf 78}, 035416 (2008).
\bibitem{marcus}J. R. Williams et al. Phys. Rev. {\bf B 80}, 045408 (2009). 
 \bibitem{sdh} I.M. Lifshitz and A.M. Kosevich, 
Sov. Phys. JETP {\bf 2}, 636 (1956), P.T. Coleridge, R. Stoner and R.Fletcher, Phys. Rev. B{\bf 39}, 195412 (1989). 
\bibitem{mass} 
This yields the BL's effective mass, which  \DM{agrees with} the 
 theoretical value \DM{of} $ 2 \hbar^2 t_{\perp}/(9 a^2 t_{\parallel}^2)$\DM{,}  where $t_{\parallel}$
 \DM{and} $t_{\perp}$ are the  in-plane
 \DM{and} transverse hopping energies\DM{,}  \cite{revueguinea}.
\bibitem{ando}T. Ando and Y. Uemura, J. Phys. Soc. Jpn. {\bf 36}, 959 (1974). 
\bibitem{sarma08}E. H. Hwang and S. Das Sarma, Phys. Rev. B{\bf 77}, 195412 (2008).
\bibitem{adhikari} S. K. Adhikari, Am. J. Phys. {\bf 54}, 362 (1986).
\bibitem{novikov}D. S. Novikov, Phys. Rev. B{\bf 76}, 245435 (2007).
\bibitem{basko} D. Basko, Phys. Rev. B {\bf 78},115432 (2008).
\bibitem{zettl}J. C. Meyer et al., 
Nano Lett., {\bf 8}, 3582 (2008).
\bibitem{katsnelson08}M. I. Katsnelson, M. A. Vozmediano and F.Guinea, Phys. Rev. B {\bf 77}, 075422 (2008).
\bibitem{Suppl} On line supplementary materials.
\end{thebibliography}
\end{document}